\documentclass[aps,pra,epsfig,twocolumn]{revtex4}
\usepackage{graphicx}% Include figure files
\usepackage{dcolumn} % Align table columns on decimal point
\usepackage{bm}
\begin{document}

\draft

\title{Qubit Assisted Probing of Coherence Between States of a Macroscopic Apparatus}

\author{Sougato Bose}

\affiliation{Institute for Quantum Information, MC 107-81,
California Institute of Technology, Pasadena, CA 91125-8100, USA\\
Department of Physics and Astronomy, University College London,
Gower St., London WC1E 6BT, UK}

\date{\today}

\begin{abstract}
I present a general scheme through which the evidence of a
superposition involving distinct classical-like states of a
macroscopic system can be probed. The scheme relies on a qubit
being coupled to a macroscopic harmonic oscillator in such a way
that it can be used to both prepare and probe a macroscopic
superposition. Two potentially realizable implementations, one
with a flux qubit coupled to a LC circuit, and the other with an
ion-trap qubit coupled to the collective motion of several ions,
are proposed.

\end{abstract}
 \maketitle

%\begin{multicols}{2}
Superpositions between distinct states of microscopic systems have
been observed in several experiments. Macroscopic systems, on the
other hand, are almost always found to be in states which are
close approximations to classical points in phase space
\cite{pointer1}. Due to environmental effects
\cite{zurekpt,caldeira}, it is very hard to observe evidences of
superpositions between such classical states of a macroscopic
system (a system with a large value of mass or equivalent).
Despite this, there has been a steady progress in schemes for
observing quantum superpositions between classical states of
macroscopic systems
\cite{leggett,zeilinger,josexpt,bose,blencowe,mancini,simon}. Some
of these are actual experiments \cite{zeilinger,josexpt} or
strongly aimed towards actual experiments \cite{blencowe,simon}.
In this letter, I will present a generic scheme in which {\em any}
qubit (assumed to remain coherent during the duration of the
scheme), when coupled appropriately to a macroscopic (material or
non-material) harmonic oscillator, can be used to both create and
probe superpositions of states of the qubit-oscillator system
which involve distinct classical states of the oscillator. This
presents a mathematically {\em unified} setting for some earlier
proposals (\cite{bose,blencowe}). Furthermore, it helps to suggest
a variety of schemes using specific qubit-harmonic oscillator
combinations. I will illustrate this diversity of the scheme by
presenting two potential implementations distinct from those
proposed before \cite{bose,blencowe}.

   There are models of decoherence
of a superposition of states of macroscopic oscillator
\cite{zurekpt,caldeira,venugopalan}, which give decoherence rates
in terms of macroscopic variables such as the mass $m$,
dissipation constant $\gamma$ and frequency $\omega$ of the
oscillator and temperature $\theta$ of the oscillator and its
environment. The domain of validity of such simplistic
(thermodynamically flavored) models, remain to be fully tested.
For a superposition of two coherent states of the oscillator
separated by a distance $\Delta x$ these models give the following
heuristic formula for the decoherence time-scale
\begin{equation}
\tau_D=\hbar^2/\{D(\Delta x)^2\}
\end{equation}
with $D=2m\gamma \hbar \omega(\bar{n} + \frac{1}{2})$ in which
$\bar{n}=[ \exp{ (\hbar \omega / k_{B}\theta)} - 1 ]^{-1}$ is the
expected number of quanta in a harmonic oscillator. For $\theta=0$
we can rewrite $\tau_D=1/(2\gamma|\Delta\alpha|^2)$ (where
$\Delta\alpha$ is the amplitude difference of the superposed
coherent states). This has been experimentally verified to hold
exactly for a single trapped ion \cite{wineland}. Our scheme, as I
will show, can be used to test the same models for higher $\theta$
and larger $m$. This will extend the work of earlier decoherence
experiments \cite{wineland,haroche} which have only tested the
dependence of $\tau_D$ on $\Delta \alpha$ and $\gamma$, while
leaving out the dependence on $m$ and $\theta$.

   I will adopt
the formula for the decoherence rate from a rigorous treatment of
the measurement of a qubit by a harmonic oscillator
\cite{venugopalan}. In this sense, the current paper has a
secondary objective of improving on earlier heuristic treatments
of decoherence in the same setting \cite{bose}. Moreover, the
scheme presents a simple application of a {\em single qubit}. As
large scale quantum computation is still somewhat distant, any use
for a single qubit is currently of interest. The macroscopic
harmonic oscillator can even be a system which usually acts as an
apparatus for the qubit. Simply setting it to work in a regime
different from that used in a typical measurement suffices to
create the conditions of our experiment. In the rest of the paper,
I will often refer to the harmonic oscillator as the apparatus.

    Interference experiments are generally done by splitting
the wave-function of a system, applying a relative phase between
the two split components and subsequently interfering these
components. For a macroscopic system, however, (1) coherent
wave-function splitters are not readily available (notable
exception being macro-molecules \cite{zeilinger}), (2) decoherence
can be unacceptably large, and (3) the initial state is mixed
(thermal). We describe below how to circumvent each problem. A
natural way to avoid the absence of a wave-function splitter for a
macroscopic system is to involve an auxiliary microscopic system
(such as a qubit) and use it to both create and probe
superpositions between distinct states of the combined microscopic
and macroscopic system. I describe the general procedure in two
steps. For convenient description, I will first assume that no
environment interacts with either the qubit or the apparatus (I
will incorporate such interactions in the next paragraph). In the
first step, a coherent superposition
$|\psi(0)\rangle_Q=(1/\sqrt{2})(|0\rangle_Q+|1\rangle_Q)$ of the
states of a qubit is prepared and a macroscopic system (which we
call an apparatus) initially in a state $|\alpha\rangle_A$ (a
close approximation to a classical point in phase space) is
allowed to interact with it. The qubit-apparatus coupling is
assumed to be such that in a time $t$ the following evolution
takes place
\begin{equation}
|\psi(0)\rangle_Q |\alpha\rangle_A \rightarrow
(1/\sqrt{2})(|0\rangle_Q|\alpha_0\rangle_A+|1\rangle_Q|\alpha_1\rangle_A).
\label{ev1}
\end{equation}
We assume that states $|\alpha_0\rangle_A$ and
$|\alpha_1\rangle_A$ are each close approximations to points in
classical phase space with coordinates being uniquely determined
by $\alpha_0$ and $\alpha_1$ respectively
($\alpha_0\neq\alpha_1$). Now an external field is applied to the
apparatus to evolve the state in the right hand side of
Eq.(\ref{ev1}) to
$\frac{1}{\sqrt{2}}(|0\rangle_Q|\alpha_0\rangle_A+e^{i\phi_{01}}|1\rangle_Q|\alpha_1\rangle_A)$.
It is ensured that the appearance of the relative phase
$\phi_{01}$ is correlated to the presence of different apparatus
states $|\alpha_0\rangle_A$ and $|\alpha_1\rangle_A$ in each
component of the above superposition. Detection of $\phi_{01}$ is
then the evidence of coherence between the states
$|0\rangle_Q|\alpha_0\rangle_A$ and
$|1\rangle_Q|\alpha_1\rangle_A$ after the first step. To detect
this phase, we take the second step in which the qubit and
apparatus are allowed to interact for another time interval $t$
such that
\begin{equation}
\frac{|0\rangle_Q|\alpha_0\rangle_A+e^{i\phi_{01}}|1\rangle_Q|\alpha_1\rangle_A}{\sqrt{2}}\rightarrow
\frac{|0\rangle_Q+e^{i\phi_{01}}|1\rangle_Q}{\sqrt{2}}|\alpha\rangle_A.
\label{ev2}
\end{equation}
In the above evolution, the apparatus is brought back to its
original state and the relative phase between the superposed
components of the joint system has become a relative phase between
the states $|0\rangle_Q$ and $|1\rangle_Q$ of the qubit. The
experiment is now concluded by determining the relative phase
$\phi_{01}$ through a measurement of the state of the qubit only.
(the same technique has been used for a trapped ion in
Ref.\cite{wineland}).

    For a macroscopic apparatus, there will also be decoherence during the evolution of $|\alpha\rangle_A$ to
$|\alpha_0\rangle_A$ and $|\alpha_1\rangle_A$ and back. We will
assume that only the states of the macroscopic apparatus decohere,
but the states of the qubit do not (at least over the time-scale
of our experiment). If the apparatus is under-damped ({\em i.e.},
it loses hardly any energy during the experiment), the evolution
in step one is modified to
\begin{eqnarray}
&&|\psi(0)\rangle\langle\psi(0)|_Q |\alpha\rangle\langle\alpha|_A
\rightarrow (1/2)(|0\rangle\langle
0|_Q|\alpha_0\rangle\langle\alpha_0|_A
\nonumber\\&+&|1\rangle\langle 1|_Q
|\alpha_1\rangle\langle\alpha_1|_A)+(e^{-{\cal
D}_{01}}/2)(|0\rangle\langle 1|_Q
|\alpha_0\rangle\langle\alpha_1|_A\nonumber\\&+&|1\rangle\langle
0|_Q|\alpha_1\rangle\langle\alpha_0|_A), \label{ev3}
\end{eqnarray}
where $e^{-{\cal D}_{01}}$ is the decoherence term (${\cal
D}_{01}>0$). Note that in Eq.(\ref{ev3}), the appearance of the
decoherence term is in one to one correspondence with the
appearance of terms $|0\rangle\langle
1|_Q|\alpha_0\rangle\langle\alpha_1|_A$ and $|1\rangle\langle
0|_Q|\alpha_1\rangle\langle\alpha_0|_A$. Thus a measurement of
this term suffices to demonstrate the partial coherence (as
quantified by $e^{-{\cal D}_{01}}$) between
$|0\rangle_Q|\alpha_0\rangle_A$ and
$|1\rangle_Q|\alpha_1\rangle_A$ after step one. We can thus
dispense with the application of the relative phase $\phi_{01}$
($e^{-{\cal D}_{01}}$ can itself be thought as arising from random
relative phases) and directly proceed to step two, in which the
modified evolution is
\begin{eqnarray}
&&(1/2)(|0\rangle\langle 0|_Q|\alpha_0\rangle\langle\alpha_0|
+|1\rangle\langle 1|_Q
|\alpha_1\rangle\langle\alpha_1|_A)\nonumber\\&+&(e^{-{\cal
D}_{01}}/2)(|0\rangle\langle 1|_Q
|\alpha_0\rangle\langle\alpha_1|_A\nonumber\\&+&|1\rangle\langle
0|_Q|\alpha_1\rangle\langle\alpha_0|_A) \rightarrow
(1/2)(|0\rangle\langle 0|_Q +|1\rangle\langle
1|_Q)\nonumber\\&+&(e^{-{\cal D}_{01}}/2)(|0\rangle\langle
1|_Q+|1\rangle\langle 0|_Q)|\alpha\rangle\langle\alpha|_A.
\label{ev4}
\end{eqnarray}
In the above, equal decoherence factors have been assumed in both
the steps. After the second step, the state of the qubit is
measured to determine ${\cal D}_{01}$ and thereby the degree of
coherence between $|0\rangle_Q|\alpha_0\rangle_A$ and
$|1\rangle_Q|\alpha_1\rangle_A$ after step one. One of the
problems with a macroscopic apparatus is that ${\cal D}_{01}$ can
be so large that the measured degree of coherence is too small.
However, in most reasonable models of decoherence, ${\cal D}_{01}$
also depends on the absolute value of the difference
$\alpha_0-\alpha_1$ and can thereby be reduced by making this
difference small. So no matter how macroscopic the apparatus is,
we can always bring ${\cal D}_{01}\sim 1$, so that some coherence
between $|0\rangle_Q|\alpha_0\rangle_A$ and
$|1\rangle_Q|\alpha_1\rangle_A$ persists. Measuring ${\cal
D}_{01}$ not only provides evidence of the partially coherent
superposition between $|0\rangle_Q|\alpha_0\rangle_A$ and
$|1\rangle_Q|\alpha_1\rangle_A$, but also allows us to
systematically test the models of decoherence of the macroscopic
apparatus.

      A third problem associated with an interference experiment
with a macroscopic system is its initial thermal state. For this,
it is necessary to assume that {\em irrespective} of the initial
state $|\alpha\rangle_A$, the same $\alpha_1-\alpha_0$ ({\em
i.e.,} the same ${\cal D}_{01}$) results. Moreover, we need to
assume that the macroscopic system is a harmonic oscillator with
$|\alpha\rangle_A$ being coherent states. Then any thermal state
can be written as $\int p(\alpha)|\alpha\rangle\langle\alpha|_A
d^2\alpha$, where $p(\alpha)$ are probabilities, which is a
mixture of coherent states $|\alpha\rangle_A$. For each component
$|\alpha\rangle\langle\alpha|_A$ of the mixture, the dynamics
described in Eqs.(\ref{ev3}) and (\ref{ev4}) would hold, and if we
measure the state of the qubit after the third step, we would
measure the same decoherence factor $e^{-2{\cal D}_{01}}$ and
thereby the degree of coherence between
$|0\rangle_Q|\alpha_0\rangle_A$ and
$|1\rangle_Q|\alpha_1\rangle_A$ at the end of the first step.

\begin{figure}
\begin{center}
\includegraphics[width=3.5in, clip]{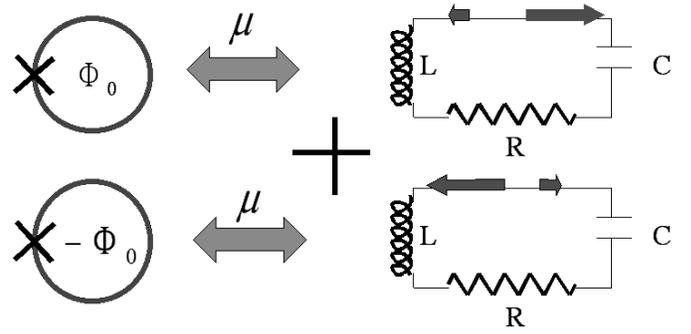}
 \caption{The figure shows the probing of macroscopic coherence by coupling a flux qubit to a
 macroscopic LC tank circuit. Being initially prepared in the state $(1/\sqrt{2})(|0\rangle_Q+|1\rangle_Q)$, the flux
 qubit induces different flux values (current directions) in the circuit corresponding to its different flux states; the resulting
 macroscopic superposition is shown in the figure.}
\label{squid1}
\end{center}
\end{figure}

  We will now propose a scheme using an explicit Hamiltonian for the qubit-apparatus-environment system, in which
the apparatus is modeled as a harmonic oscillator. The first and
the second steps will now be two successive parts of the same time
evolution. We will use a model studied by Venugopalan
\cite{venugopalan} in the context of a quantum measurement
(although, we will use it in a parameter regime different from
that of a measurement). The Hamiltonian for the model can be
written as
\begin{equation}
H= H_{Q}+H_{A}+H_{QA}+H_{AE}, \label{ham1}
\end{equation}
where $H_Q$ is the qubit Hamiltonian, $H_{QA}$ is the
qubit-apparatus interaction, $H_A$ is the apparatus Hamiltonian
and $H_{AE}$ is the apparatus-environment interaction. These terms
are explicitly given by
\begin{eqnarray}
H_{A}=\frac{p^{2}}{2m} + \frac{1}{2}m \omega^{2} x^{2},~
 H_{Q}=\lambda \sigma_{Z},~
H_{QA}= \epsilon x
\sigma_{Z} \\
H_{AE}=\sum_{j}  \frac{P_{j}^{2}}{2M_{j}} + \frac{M_{j}
\Omega_{j}^{2}}{2} \bigl ( X_{j}-\frac{C_{j}x}{M_{j}
\Omega_{j}^{2}} \bigr )^{2}. \label{ham2}
\end{eqnarray}
Here $x$ and $p$ denote the position and momentum of the harmonic
oscillator (apparatus), $\lambda \sigma_{Z}$ is the the
Hamiltonian of the qubit (for simplicity, we will set $\lambda=0$
for the rest of the paper; this does not affect our conclusions),
and $\epsilon$ is the strength of the qubit-apparatus coupling.
The last term represents the Hamiltonian for the environment (a
bath of oscillators) and the apparatus-environment interaction.
$X_{j}$ and $ P_{j}$ are the position and momentum coordinates of
the jth harmonic oscillator of the bath, $C_{j}$'s are the
coupling strengths and $\Omega_{j}$s are the frequencies of the
oscillators comprising the bath \cite{venugopalan}.

     Venugopalan \cite{venugopalan} has found the complete solutions to
the model for all parameter domains of the Hamiltonian given by
Eqs.(\ref{ham1}) and (\ref{ham2}) in the Markovian limit
($k_{B}\theta<<\hbar\Omega_{\mbox{cut}} $ where $k_{B}=$ Boltzmann
constant, $\hbar=$ Plank constant and $\Omega_{\mbox{cut}}=$
cutoff frequency of the environmental oscillators). Subsequently,
Venugopalan has used the solutions only in the case of an
overdamped oscillator ($\omega << \gamma$ where where $\gamma$ is
the relaxation rate of the oscillator) in the long time
($t\rightarrow\infty$) limit to analyze a measurement of the
qubit's state by the oscillator. Our aim here is to investigate
coherence between distinct states of the oscillator and not to
complete a measurement (in course of which, all coherence will be
lost). Hence we concentrate on the {\em under-damped} regime of
parameters of the oscillator ($\gamma << \omega$) for times $t\leq
T=2\pi/\omega$ ({\em i.e.,} $T$ is one oscillation period of the
oscillator). In this parameter domain, if the oscillator starts in
an initial coherent state
($\rho_A(0)=|\alpha\rangle\langle\alpha|_A$) and the qubit in the
state $(1/\sqrt{2})(|0\rangle_Q+|1\rangle_Q)$ then the evolution
of the system from $t=0$ to $t=T/2$ is taken to be step one of our
protocol. At the end of this step the state is given by
\begin{eqnarray}
&&\rho_{QA}(T/2)= \frac{1}{2}(|0\rangle\langle
0|_Q|\alpha_0\rangle\langle\alpha_0|_A +|1\rangle\langle 1|_Q
|\alpha_1\rangle\langle\alpha_1|_A)\nonumber\\&+&\frac{e^{-{\cal
D}_{01}}}{2}(e^{-i\phi_{01}}|0\rangle\langle 1|_Q
|\alpha_0^{'}\rangle\langle\alpha_1^{'}|_A+e^{i\phi_{01}}|1\rangle\langle
0|_Q|\alpha_1^{'}\rangle\langle\alpha_0^{'}|_A)\nonumber\\
\label{statepi}
\end{eqnarray}
where $|\alpha_j\rangle_A$ and $|\alpha_j^{'}\rangle_A$ (with
$j=0,1$) are coherent states of the apparatus with
\begin{equation}
\alpha_j =-\alpha+(-1)^j \Delta x/2\delta_x,~
\alpha_j^{'}=\alpha_j+i \Delta p/2\delta_p.
\end{equation}
where $\delta_x=\sqrt{\hbar/2m\omega}$ and
$\delta_p=\sqrt{m\hbar\omega/2}$ are the coherent state position
and momentum spreads, $\Delta x=2\epsilon/m\omega^2$ is the
separation between the coherent states $|\alpha_0\rangle_A$ and
$|\alpha_1\rangle_A$,$\Delta p=D\Delta x/2\omega\hbar^2$ is an
unwanted momentum shift, ${\cal D}_{01}=\frac{D(\Delta
x)^2}{\hbar^2}\frac{T}{2}$ is the decoherence exponent and
$\phi_{01}=-4\pi{\cal D}_{01}$. If we define a scaled
dimensionless qubit-apparatus coupling strength $\chi$ as
$\epsilon\delta_x/\hbar\omega$ and the quality factor $Q$ of the
oscillator as $Q=\omega/\gamma$, then (in the high temperature
$\bar{n}\approx k\theta/\hbar\omega>>1$ limit) we can rewrite the
orders of magnitudes of the earlier quantities as
\begin{equation}
\frac{\Delta x}{2\delta_x}\sim\chi,~\frac{\Delta
p}{2\delta_p}\sim\frac{\chi\bar{n}}{Q},~{\cal
D}_{01}\sim\frac{\chi^2\bar{n}}{Q}. \label{scaled}
\end{equation}
We want Eq.(\ref{statepi}) to correspond to the state in the right
hand side of Eq.(\ref{ev3}) (apart from the unimportant phase
factor $\phi_{01}$) and thus require $\alpha_j\approx
\alpha_j^{'}$, which in turn implies $\exp{[-\{\frac{\Delta
p}{2\delta_p}\}^2]}=\exp{[-\{\frac{\chi\bar{n}}{Q}\}^2]}\approx
1$. This is clearly satisfied if we choose our parameters such
that
\begin{equation}
Q^2>>(\chi\bar{n})^2. \label{cond1}
\end{equation}
We would also want (ideally) $\Delta x$ to be larger than or
comparable to a coherent state width. This implies
\begin{equation}
\chi \geq 1. \label{cond2}
\end{equation}
Moreover, we want some coherence to be present between
$|0\rangle_Q|\alpha_0\rangle_A$ and
$|1\rangle_Q|\alpha_1\rangle_A$ in $\rho_{QA}(T/2)$. We thus want
${\cal D}_{01}\sim 1$ which implies
\begin{equation}
Q \sim \chi^2\bar{n}.\label{cond3}
\end{equation}
When the qubit-apparatus system is allowed to evolve further from
$t=T/2$ to $t=T$ (this is the step two of our protocol), the final
state is given by
\begin{eqnarray}
\rho_{QA}(T)&=&(1/2)\{|0\rangle\langle 0|_Q +|1\rangle\langle
1|_Q\nonumber\\&+&e^{-2{\cal D}_{01}}(|0\rangle\langle
1|_Q+|1\rangle\langle
0|_Q)\}\otimes|\alpha\rangle\langle\alpha|_A.
\end{eqnarray}
Detecting the decoherence factor $e^{-2{\cal D}_{01}}$ by
measuring the state of the qubit now corresponds to measuring a
signature of the partial coherence between
$|0\rangle_Q|\alpha_0\rangle_A$ and
$|1\rangle_Q|\alpha_1\rangle_A$ at $t=T/2$.

  A further aim of our scheme is to test the
dependence of ${\cal D}_{01}$ on $\theta$ and $m$. From the
expression of ${\cal D}_{01}$ in Eq.(\ref{scaled}), and
$\bar{n}\propto \theta$, it follows that ${\cal D}_{01}$ will vary
in direct proportion with $\theta$. The dependence on $m$ is
trickier to test. One has to create the {\em same} $\Delta x$ for
different $m$ to make a fair study of the dependence of ${\cal
D}_{01}$ with $m$. As Eq.(\ref{scaled}) suggests, ${\cal
D}_{01}\propto\chi^2\sim (\Delta x)^2/4\delta_x^2\propto m (\Delta
x)^2$. Thus if one creates the same $\Delta x$ for different $m$,
then ${\cal D}_{01}$ should be found to be directly proportion to
$m$.

  We now proceed to suggest some potentially realizable experiments in which the
above scheme can be tested. In the first realization, we take the
qubit to be a {\em flux qubit} \cite{Makhlin} coupled by a
flux-flux coupling to a LC tank circuit \cite{prance}, which is
the macroscopic harmonic oscillator (apparatus) as shown in
Fig.\ref{squid1}. The $|0\rangle_Q$ and $|1\rangle_Q$ states of
the qubit correspond to flux values $\pm\Phi_0$ with
$\Phi_0=\hbar/2e$ (where $e$ is electronic charge). The flux
variable of the qubit can thus be written as the operator $\Phi_0
\sigma_z$. For the qubit-circuit coupled system \cite{everitt}
\begin{equation}
H_{A}=(C/2)\dot{\Phi}^2+(1/2L)\Phi^2,~H_{QA}=\mu \Phi
(\Phi_0\sigma_z)/\Lambda \label{squidH}
\end{equation}
where $C$ and $L$ are the capacitance and the inductance of the
circuit, $\Phi$ is the flux through the circuit, $\Lambda$ is the
inductance of the qubit, $\mu$ is a dimensionless coupling (flux
linkage) between the qubit and the circuit. From
Eqs.(\ref{squidH}) it is evident that $\Phi$ and $C$ are the
electrical counterparts of $x$ and $m$ respectively. For this
system,
$\epsilon=\mu\Phi_0/\Lambda,~\gamma=R/2L,~\omega=1/\sqrt{LC}$.

   We now make an explicit choice of parameters of the
qubit and the circuit to show that the three conditions for our
experiment can be satisfied. We choose $\Lambda\sim 100$pH
\cite{Makhlin} for the flux qubit. For the circuit, we choose
$L=\sim100$$\mu$H, $C\sim100$pF (a macroscopic value in comparison
to the available $fF$ capacitances \cite{Makhlin}), $R\sim
1\Omega$. This gives $Q\sim 10^3$ at $\omega\sim 10$MHz ($Q\sim
10^6$ for $\omega\sim 10$KHz \cite{grav} to $Q\sim 10^2$ for
$\omega\sim 1$GHz \cite{cmos} are achievable). We assume the
temperature of the circuit and its environment to be $\theta\sim
10$mK (same as that of the flux qubit \cite{Makhlin}). With this
choice, $\bar{n}\sim 100$ and
\begin{equation}
Q^2/\bar{n}^2\sim 10^2,
\end{equation}
thereby satisfying Eq.(\ref{cond1}). We choose $\mu\sim 10^{-6}$
to have
\begin{equation}
\chi \sim \sqrt{10},~\chi^2 \bar{n}\sim 10^3 \sim Q
\end{equation}
which implies that Eqs.(\ref{cond2}) and (\ref{cond3}) are
satisfied.

\begin{figure}
\begin{center}
\includegraphics[width=3in, clip]{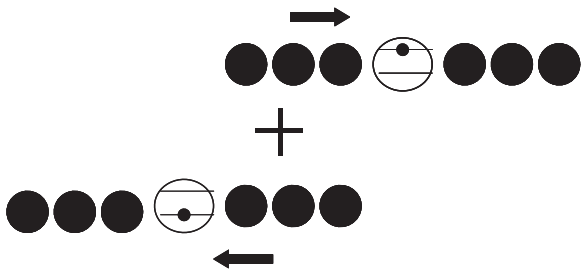}
 \caption{The figure shows the probing of macroscopic coherence by coupling the internal levels of a single ion to
the collective motional state of several ions in a trap. Being
initially prepared in the state
$(1/\sqrt{2})(|0\rangle_Q+|1\rangle_Q)$, the
 internal level qubit displaces the collective motional state differently corresponding to its two states; the resulting
 macroscopic superposition is shown in the figure.}
\label{ion}
\end{center}
\end{figure}

    Next we propose a different physical realization using a linear ion trap \cite{wineland}. In this case,
we use the internal levels of a single trapped ion as the qubit
and the collective motion of $N$ ions (with large $N$) as the
macroscopic oscillator as shown in Fig.\ref{ion}. The
qubit-apparatus coupling term of the Hamiltonian for the system is
\cite{wineland3}
\begin{equation}
H_{QA}=(\hbar g\eta_0/\sqrt{N}\delta_x)x\sigma_z,
\end{equation}
where $g$ is the vacuum Rabi frequency and $\eta_0$ is the
Lamb-Dicke parameter for a single ion. We assume $N\sim 100$,
$\omega\sim 1$MHz (an order less than in \cite{blatt}), $g\sim
100\sqrt{10}$MHz (300 times higher than in \cite{wineland4}) and
$\eta_0\sim 0.1\sqrt{10}$ (nearly same as in \cite{wineland4}).
With these choices
\begin{equation}
\chi\sim g\eta_0/\sqrt{N}\omega \sim 10,
\end{equation}
which satisfies Eq.(\ref{cond2}). We now choose $\theta\sim
0.1$mK, so that $\bar{n}\sim 10$. We also note that for the
natural (ambient) reservior $\gamma \sim 1$KHz \cite{wineland4},
which implies a quality factor $Q\sim 10^3$. This implies
\begin{equation}
(Q/\chi \bar{n})^2 \sim 10^2,~\bar{n}\chi^2/Q\sim 1,
\end{equation}
which satisfies Eq.(\ref{cond1}) and Eq.(\ref{cond3}).

    In this paper, I have presented a general scheme for
probing evidences of superpositions which involve distinct
classical-like states of a macroscopic object. Though realizations
with two specific qubit-harmonic oscillator combinations are
proposed, the paper opens up the scope for applying to any other
combination. It also provides a unified mathematical setting for
some earlier proposals \cite{bose,blencowe} and can be used to
probe the dependence of decoherence rate on the mass and
temperature. In comparison to the schemes used in decoherence
experiments so far \cite{haroche,wineland}, this is more easily
extendable to macroscopic oscillators where it might be difficult
to switch qubit-oscillator interactions on and off (such as the
radiation pressure interaction \cite{bose,mancini}) and where the
superposition itself forms over such a length of time that
decoherence during the formation is important.

  This work is supported by the NSF under Grant Number
   EIA-00860368. I thank P. Delsing for valuable comments.

%\end{multicols}


\begin{thebibliography}{99}
\bibitem{pointer1}
W. H. Zurek, S. Habib and J. P. Paz, Phys. Rev. Lett. {\bf 70},
1187 (1993).
\bibitem{zurekpt}
W. H. Zurek, Physics Today {\bf 44} (10), 36 (1991).
\bibitem{caldeira}
A. O. Caldeira and A. J. Leggett, Physica A {\bf 121}, 587 (1983).
\bibitem{leggett}
A. J. Leggett and A. Garg, Phys. Rev. Lett. {\bf 54}, 857 (1985).
\bibitem{zeilinger}
M. Arndt {\em et. al.}, Nature {\bf 401}, 680 (1999).
\bibitem{josexpt}
J. Friedman {\em et. al.} Nature {\bf 406} 43 (2000); C. H. van
der Wall {\em et. al} Science {\bf 290} 773 (2000).
\bibitem{bose}
S. Bose, K. Jacobs and P. L. Knight, Phys. Rev. A {\bf 59}, 3204
(1999).
\bibitem{blencowe}
A. D. Armour, M. P. Blencowe and K. C. Schwab, Phys. Rev. lett
{\bf 88}, 148301 (2002).
\bibitem{mancini}
S. Mancini, V. Giovannetti, D. Vitali, and P. Tombesi, Phys. Rev.
Lett. {\bf 88}, 120401 (2002).
\bibitem{simon}
W. Marshall, C. Simon, R. Penrose and D. Bouwmeester,
quant-ph/0210001 (2002).
\bibitem{venugopalan}
A. Venugopalan, Phys. Rev. A {\bf 61}, 012102 (1999).
\bibitem{wineland}
C. J. Myatt {\em et. al.}, Nature {\bf 403}, 269 (2000).
\bibitem{haroche}
M. Brune {\em et. al.}, Phys. Rev. Lett. {\bf 77} 4887 (1996).
\bibitem{Makhlin}
Y. Makhlin, G. Sch\"{o}n and A. Shnirman, Rev. Mod. Phys. {\bf
73}, 357 (2001).
\bibitem{prance}
R. J. Prance {\em et. al.},  Phys. Rev. Lett. {\bf 82}, 5401
(1999).
\bibitem{everitt}
M. J. Everitt {\em et. al.}, Phys. Rev. B {\bf 63}, 144530 (2001).
\bibitem{grav}
M. Bonaldi {\em et. al.}, Rev. of Sci. Inst. {\bf 70}, 1851
(1999).
\bibitem{cmos}
P. Kinget and R. Frye, Proc. of the ESSCIRC, pp. 364-367 (1998).
\bibitem{wineland3}
D. J. Wineland {\em et. al.}, J. Res. Natl. Inst. Stand. Technol.
{\bf 103}, 259 (1998).
\bibitem{blatt}
F. Schmidt-Kaler {\em et. al.},Nature {\bf 422} (2003).
\bibitem{wineland4} C. Monroe {\em et. al.}, Science {\bf 272},
1131 (1996).




\end{thebibliography}
\end{document}